\documentclass[runningheads]{svmult}

\usepackage{makeidx}   % allows index generation
\usepackage{graphicx}  % standard LaTeX graphics tool
\usepackage{subeqnar}  % subnumbers individual equations
\usepackage{multicol}  % used for the two-column index
\usepackage{physprbb}  % modified textarea for proceedings,
\makeindex             % used for the subject index
\begin{document}
\title*{Chemical Evolution in the Carina dSph}
\toctitle{Chemical Evolution in the Carina Dwarf Spheroidal}
\titlerunning{Chemical Evolution in Carina}
\author{A. Koch\inst{1} \and E.K. Grebel\inst{1} \and D. Harbeck\inst{2} \and M. I. Wilkinson\inst{3} \and 
J. T. Kleyna\inst{3,4}\and  \\ G. F. Gilmore\inst{3} \and  R.F.G. Wyse\inst{5} \and N. W. Evans\inst{3}}
\authorrunning{A. Koch et al.}
\institute{Astronomical Institute of the University of Basel, CH-4102 Binningen, Switzerland
\and Department of Astronomy, University of Wisconsin, Madison, WI 53706, USA
\and  Institute of Astronomy, Cambridge University, Cambridge CB3 0HA, UK
\and  Institute for Astronomy, University of Hawaii, Honolulu, HI 96822, USA
\and  The Johns Hopkins University, Baltimore, MD 21218, USA  
}
\maketitle              
\begin{abstract}
We present metallicities for 487 red giants in the Carina dwarf spheroidal (dSph) galaxy that were obtained from 
FLAMES low-resolution Ca triplet (CaT) spectroscopy. 
We find a mean [Fe/H] of $-1.91$\,dex
with an intrinsic dispersion of 0.25\,dex, whereas the full spread in metallicities is at least one dex.
The analysis of the radial distribution of metallicities reveals that an excess of metal poor stars resides in a region of
larger axis distances.
These results can constrain evolutionary models and are discussed in the context of chemical evolution in the Carina dSph. 
\end{abstract}
\section{Introduction}
Analyses of the faint Carina dSph have revealed that it contains a %n intriguing
variety of stellar populations (e.g.,~[4]), exhibiting prominent old ($>$11\,Gyr) and intermediate-age (5--6 and 3\,Gyr) populations. 
This implies that Carina must have undergone several star forming (SF) episodes with 
at least three significant pulses. Despite this wide spread in ages, its colour-magnitude diagram features a remarkably narrow RGB. The reason for that 
can be an age counteracting spread in metallicities, where metal rich, young stars have colours comparable to the older, more metal poor ones.
Such a possible age-metallicity degeneracy can be overcome if accurate and independent [Fe/H] measurements are obtained so that 
the remaining parameter of age can be estimated from isochrones. Moreover, the overall shape and spatial variations of the metallicity distribution function (MDF) itself contain 
valuable implications for analysing Carina's unusual SF history.
\section{Observations, Reduction and Calibration}
In the course of an ESO Large Programme, we observed 1257 red giants covering five fields in Carina out to the 
tidal radius.
Our observations were performed during 23 nights with the multi-object spectrograph FLAMES at the VLT in low-resolution mode, centered at the 
near infrared CaT.
The data were reduced using the standard FLAMES reduction pipline. 
Since sky contamination from bright emission lines is strong in the CaT region, we calculated an average sky-spectrum 
from 20 dedicated skyfibres, which then was carefully subtracted from the science spectra.
Typical signal-to-noise (S/N) ratios for our spectra lie between 25 and 
150 pixel$^{-1}$. 
After rejection of foreground stars and stars with too low a S/N, 487 radial velocity members were analysed.
Finally, metallicities were derived from the reduced equivalent width W$^{\prime}$ ([1],[5]). %
\section{The Observed and Predicted Metallicity Distributions}
The resulting MDF for our 487 red giants is shown in the top panel of Fig.~1. 
It is peaked at an average [Fe/H] of $-1.91\pm 0.01$\,dex, which is in excellent agreement with the spectroscopic study
of~[6], who found a value of $-1.99\pm 0.08$\,dex.
We find a dispersion of $\sigma$=0.25\,dex and 
as Fig.~1 implies, the full spread of [Fe/H] is at least one dex. 
This suggests that the wide range in age is in fact counteracted by a wide spread in metallicity, 
which can explain Carina's 
narrow RGB and confirms that a complex mixture of stellar populations as in Carina does not necessarily contradict the RGB's narrowness. 
The different stellar populations in Carina have also different 
spatial distributions, where, the intermediate age red clump stars are clearly concentrated 
towards the center compared to the old HB (see [1]). Thus we plot in Fig.~1 (bottom panel) the error-weighted MDFs for three different regions 
along the galaxy's axes.
While the curves for the central region and larger minor axis distances do not differ largely and also resemble those along the major axis, 
 there is a shift of the MDF's peak towards the metal poor end by 0.2\,dex when concentrating on the outer, northwestern region (lower axis distances).
Fig.~1 also overplots a model curve from~[3], showing the predicted MDF for Carina.
This model uses a SF efficiency of 0.1\,Gyr$^{-1}$ and a galactic wind efficiency at the lower end of
wind rates in dSphs.
Such a lower wind rate nicely explains the smooth decline of the observed MDF towards the metal rich end, as a higher wind rate would expel
enriched gas more efficiently thus preventing the formation of this metal rich tail. 
Secondly, the low SF rate is able to explain the metal poor peak of the MDF in terms of a moderate enrichment history.
However, remaining caveats are the higher fraction of predicted extremely metal poor stars and the lower number of metal poor stars around [Fe/H]$\approx  
 -2$, which can be understood in the context of the low infall timescale of gas assumed in dSphs ([3]).
With our accurate [Fe/H] measurements at hand we will in the next steps pursue the underlying age distribution and use the abundance ratios from 
our multi-object high-resolution spectroscopy 
to further unravel the mechanisms governing Carina's evolutionary history. 
\begin{figure}[h]
\begin{center}
\includegraphics[width=12cm]{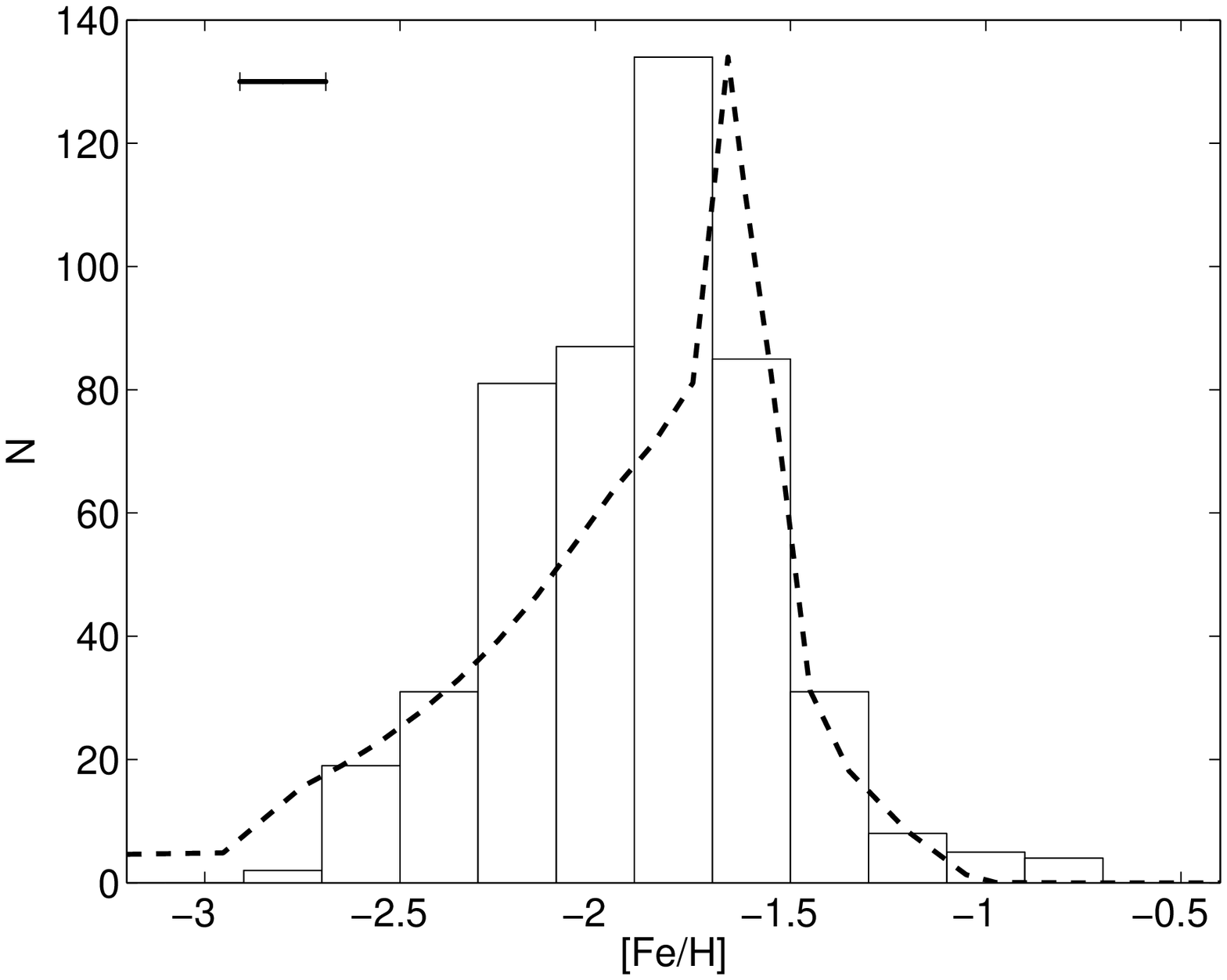}
\includegraphics[width=12cm]{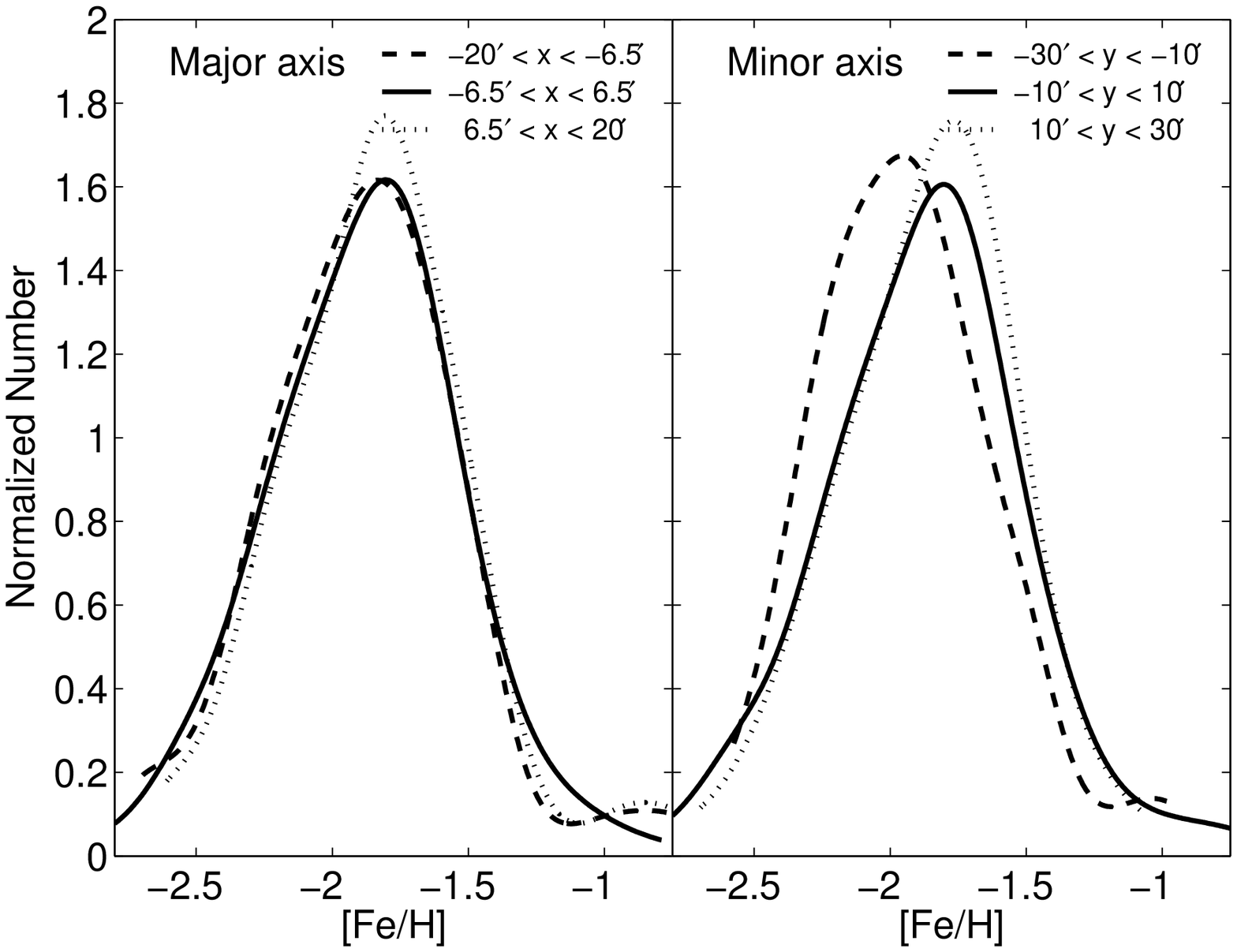}
\end{center}
\caption[]{Top panel: Metallicity distribution for our 487 red giants in Carina. The dashed line is the \lq\lq best-fit model\rq\rq \,prediction 
from [3]. The bottom plot 
shows spatially separated MDFs in three different regions of Carina, measured along the major and minor axes.}
\end{figure}


\begin{thebibliography}{6}
\addcontentsline{toc}{section}{References}
%
\bibitem{cole04} Cole, A.A., et al., MNRAS, 347, 367
%
\bibitem{harbeck01} Harbeck, D., et. al. 2001, AJ, 122, 3092  
%
\bibitem{lanfr04} Lanfrachi, G., \& Matteucci, F. 2004, MNRAS, 351, 1338
%
\bibitem{monelli03} Monelli, M., et al. 2003, AJ, 126, 218 
%
\bibitem{rutledge97} Rutledge, G.A.,  Hesser, J.E., Stetson, P.B. 1997, PASP, 109, 907 
%
\bibitem{smeckerhane99} Smecker-Hane, T.A., et al. 
 1999,  ASP Conf. Ser., 192, 159 
%
\end{thebibliography}
\end{document}